\documentclass[preprint]{elsarticle}
\usepackage{hyperref}
\usepackage[T1]{fontenc}
\usepackage[utf8]{inputenc}
\usepackage{lmodern}
\usepackage[miktex]{gnuplottex}
\usepackage{graphicx}
\usepackage{amsmath}
\usepackage{bm}
\usepackage{cleveref}
\usepackage{algorithm}
\usepackage{listings}
\usepackage{algpseudocode}
\usepackage{xfrac}
\usepackage{amssymb}

\usepackage{color}

\newcommand{\refFig}[1] {Figure \ref{#1}}

\newcommand{\refEqu}[1] {Equation \ref{#1}}
\newcommand{\refAlg}[1] {Algorithm \ref{#1}}

\newcommand{\code}[1] {
  {\textit{#1}}
}

 \newcommand{\bfE}{\mathbf{E}}
\newcommand{\bE}{\mathbf{E}}
\newcommand{\bfB}{\mathbf{B}}
 
\newcommand{\bB}{\mathbf{B}}

\newcommand{\bJ}{\mathbf{J}}

\newcommand{\bv}{\mathbf{v}}

\newcommand{\br}{\mathbf{r}}

\journal{Computer Physics Communications}

\begin{document}

\begin{frontmatter}

\title{Performance analysis and implementation details of the Energy Conserving Semi-Implicit Method code (ECsim)}

\author{Diego Gonzalez-Herrero\corref{mycorrespondingauthor}}
\cortext[mycorrespondingauthor]{Corresponding author}
\ead{diego.gonzalez@kuleuven.be}
\author{Elisabetta Boella}
\author{Giovanni Lapenta}

\address{Department of Mathematics, KU Leuven, University of Leuven, Belgium}

\begin{abstract}
We present in this work the implementation of the Energy Conserving Semi-Implicit 
Method in a parallel code called ECsim. This new code is a
three-dimensional, fully electromagnetic particle in cell (PIC) code. It is  written in C/C++ 
and uses MPI to allow massive parallelization. 
ECsim is unconditionally stable in time, eliminates the finite grid instability, has
the same cycle scheme as the explicit code with a computational cost comparable to
other semi-implicit PIC codes. All this features make it a very valuable tool 
to address situations which have not been possible to analyze until now with other PIC codes. 
In this work, we show the details of the algorithm implementation and we study
its performance in different systems. ECsim is compared with another semi-implicit PIC code
with different time and spectral resolution, showing its ability to address situations 
where other codes fail.
\end{abstract}

\begin{keyword}
Particle in cell (PIC), 
Semi-implicit particle in cell, 
Exactly energy conserving
\end{keyword}

\end{frontmatter}


\section{Introduction}

Patrice In Cell (PIC) simulations are widely used in plasma physics, mainly to study phenomena where the kinetic behavior of the
<<<<<<< HEAD
particles has an important role. The basic idea of these algorithms is to describe the plasma using the statistical distribution functions
of ions and electrons, which are sampled using computational particles. Each of these particles account for
a cloud of physical particles and interact with each other through the electromagnetic fields, which are
computed on a discrete grid \citep{hockney-eastwood, birdsall-langdon}. The key difference between different 
PIC algorithms is how the particles and the field are coupled. In the
simplest case, the fields are assumed to be static while the particles are updated (explicit PICs), and 
then the fields are updated by using the new new position and velocity of the particles. These algorithms 
are limited by the fact that the smallest scales involved in the problem need to be resolved, which limits the
range of time step and grid spacing. To overcome these limitations the PIC algorithm has evolved to the implicit
methods. In this case the particles and fields are updated together in a non-linear iterative scheme 
\citep{Chen-jcp-2011, markidis2011energy}. However these algorithms are very expensive computationally speaking. 
An intermediate approach are the semi-implicit methods \citep{Brackbill:1985}, where the coupling between the
fields and the particles is approximated through linearization, which avoids the non-linear iterations and removes some of the
limitations of the explicit methods \citep{directimplicit, brackbill-forslund}. In general semi-implicit methods do not
conserve the energy of the system, which introduces some limitations on the resolution of the simulations.
A new semi-implicit method which conserves the energy exactly has been recently published \citep{lapenta2016exactly}. The
novelty of this algorithm is that it ensures that the energy transferred between particles and fields is exactly the same without
using any non-linear iterations (its cycle scheme is similar to the one of the explicit methods). This method has proved to
be very useful to study multi-scale problems covering a wide range of resolutions \citep{Lapenta2017}.

The aim of this work is to explain in detail the implementation of this algorithm in a 3D parallel code, called ECsim, 
and to study its performance in different super computers. The critical points of the algorithm are analyzed in order to 
detect bottlenecks which could reduce its performance. Finally we test the code against a well-known situation: whistler 
waves. We compare ECsim with another semi-implicit code and we analyze the influence of the spatial and time 
resolution on the results.

\section{Energy conserving semi-implicit method}

In this section we will summarize the equations of the Energy conserving semi-implicit method which have been used in the
implementation on the code. The detailed description of the algorithm and the derivation of the equations can be found in 
\cite{lapenta2016exactly}.

The key idea of this new method is that it allows us to compute analytically the current generated by the particles in one time step without any approximations.
This causes the exchange of energy between the field and the particles to be perfectly balance, which in the end leads to a perfect energy conservation. 
The $\theta$-scheme is used for time discretization and the position is staggered half a time step with respect to the velocities and the fields.
We can split the code in three different phases: moment gathering, field solver and particle mover.

\begin{itemize}
  \item Moment gathering: From the position and velocity of the particles and with the values of the fields at this time step $n$
   we compute the implicit current $\widehat{J}_N$ and the mass matrices $M_{NN\prime}^{ij}$.
  \begin{eqnarray}
    \widehat{\bJ}_N = \frac{1}{V_N} \sum_s \sum_p q_p  \widehat{\bv}_p W_{pN},\\
    \label{EQjhat}
    M_{NN^\prime}^{ij} = \sum_s \frac{\beta_s}{V_N} \sum_p q_p {\alpha}^{ij,n}_p W_{pN^\prime} W_{pN},
    \label{EQmass}
  \end{eqnarray}
  where the summation over $p$ is extended to all the particles of a single species and the one over $s$ to all the species.
  $V_N$ is the control volume of the node $N$, $q_p$ and $\bv_p$ the charge and velocity of the particles, $W_{pN}$ the linear interpolation function
  between the particle $p$ and the node $N$ of the grid and $\beta =  q_p \Delta t/(2 m_p)$, $\Delta t$ being the time step and $m_p$ the mass
  of the particle. The ${\alpha}^{ij,n}_p$ matrix is given by the expression:
  \begin{equation}
    {\alpha}_p^n =  \frac{1}{1+(\beta_s B_p^{n})^2} \left(\mathbb{I}-\beta_s \mathbb{I} \times \bB_p^n +\beta_s^2
                    \bB_p^n \bB_p^n \right),
    \label{EQalpha}
  \end{equation}
  where $\bB_p^n$ is the magnetic field interpolated from the grid to the particle position.

  \item Field solver: Once we have computed the implicit current and the mass matrices we can calculate the electric and
  magnetic fields by solving the system:
  \begin{equation}
  \begin{array}{ccc}
    \displaystyle \nabla_C \times \bfE_N^{n+\theta} + \frac{1}{c} \frac{\bfB_C^{n+\theta}-\bfB_C^n}{\theta \Delta t} =0,\\ \\
    \displaystyle \nabla_N \times \bfB_C^{n+\theta} - \frac{1}{c} \frac{\bfE_N^{n+\theta}-\bfE_N^n}{\theta \Delta t} =\frac{4\pi}{c} \left( \widehat{\bJ}_N+\sum_{N^\prime} M_{NN^\prime}
    \bE_{v^\prime}^{n+\theta} \right),
  \end{array}
  \label{EQmaxwell}
  \end{equation}
	where the subscript $N$ indicates that the magnitude is computed on the vertices and $C$ on the cell centers.

  The fields in the next time step can be obtained using the values in the previous time step and the ones in the intermediate point $\theta$:
  \begin{eqnarray}
    \bfE_N^{n+1} &=& \frac{1}{\theta}\bfE_N^{n+\theta} - \frac{1- \theta}{\theta} \bfE_N^n ,
  \end{eqnarray}
	and the same applies for the magnetic field.

  \item Particle mover: The electric field obtained in the previous step is used to update the velocities of the particles:
  \begin{eqnarray}
    \bar{\bv}_p &=& \alpha^n \left(\bv_p^n + \beta_s \bfE_p^{n+\theta}(\br_p^{n+1/2})\right), \\
    \bv_p^{n+1} &=& 2 \bar{\bv_p} - \bv_p^n,
    \label{EQupdateV}
  \end{eqnarray}
  where $\br_p$ is the position of the particle $p$ and $\alpha$ is the same matrix used in the moment gathering. Finally the position
  is updated with the new value of the velocity:
  \begin{eqnarray}
    \br_p^{n+3/2} = \br_p^{n+1/2} + \Delta t \bv_p^{n+1},
  \end{eqnarray}

\end{itemize}

If $\theta = 0.5$ then the method is second order accurate and the energy is exactly conserved. This parameter does not affect the 
stability of the method.

\section{Description of the code}

In this section we will describe in detail the implementation of the Energy Conserving Semi-Implicit Method.
The whole project has been developed in C/C++ and uses MPI for parallel computation. The total physical domain
is divided among the processes following a Cartesian topology. Each process controls the field values in its domain 
and the particles which are "living" in it.
The code is divided in two main classes: \code{Fields} and \code{Particles}, and we can distinguish three main 
parts:
\begin{itemize}
  \item Moment gathering: In this phase the particle quantities are interpolated to the grid (where
        the fields are computed).
  \item Field Solver: In this part the linear system given in \ref{EQmaxwell} is solved.
  \item Particle Mover: Using the values of the electric and magnetic fields the velocity and 
        position of the particles are updated.
\end{itemize}

The \code{Fields} class contains all the information related to the electromagnetic fields, including the electric field
in the nodes and magnetic fields in the nodes and in the centers of the cells, the implicit current in the nodes and the mass matrices
in the nodes. Each component of the vectors is stored independently in a three-dimensional array and each element of the mass
matrix is stored in a four-dimensional array.
Apart from all these variables, several additional vectors are allocated since they are needed in the field solver. 

\begin{figure}[h]
   \center
   \includegraphics[width=0.5\columnwidth]{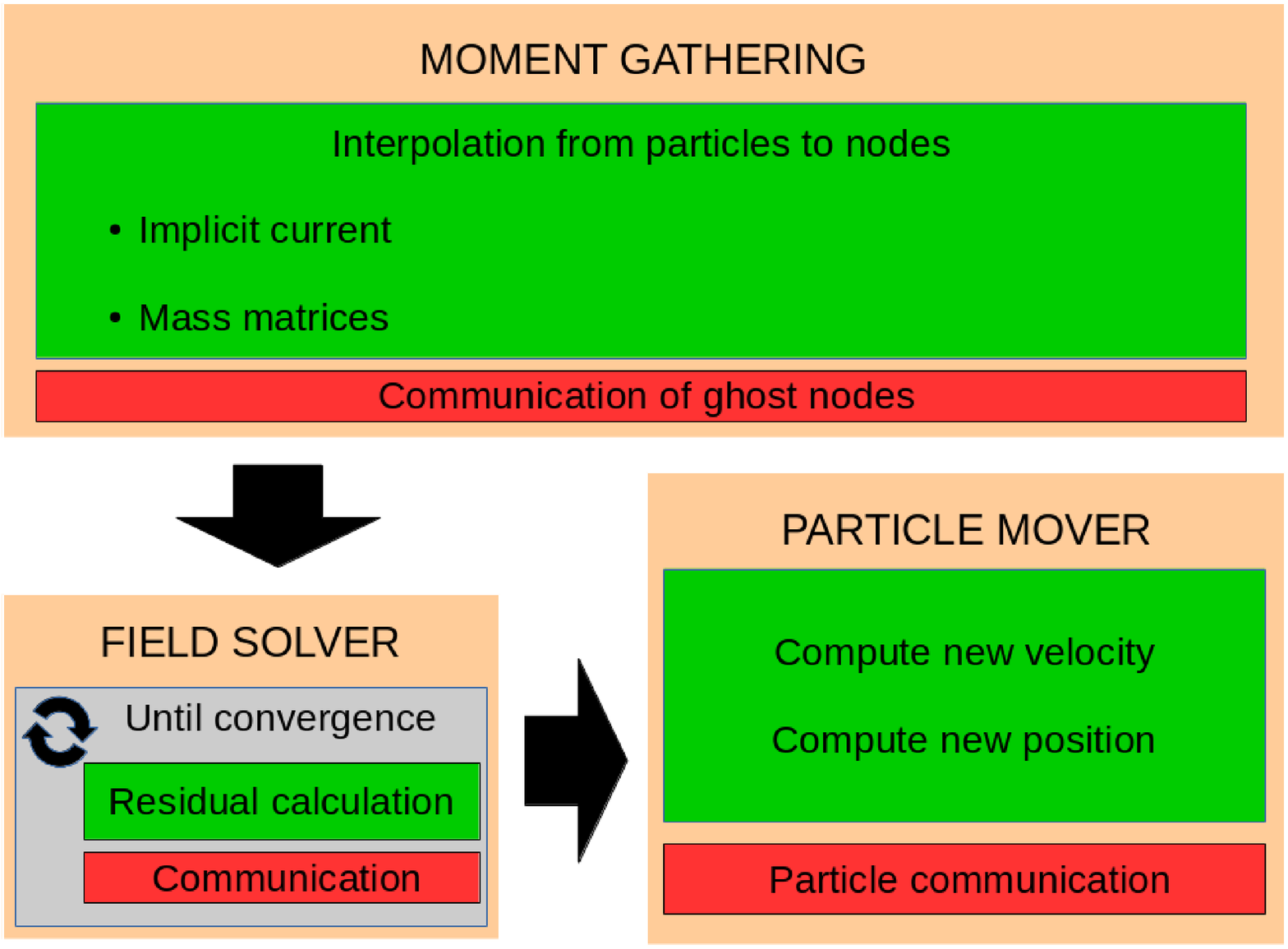}  
   \caption{Block diagram of the main loop of the code}
   \label{fig:scheme}
\end{figure}

The \code{Particles} class has all the data in relation with the particles of a single species: their position,
velocity and charge. In the main loop there will be as many instances of this class as the number of species used.
For each species, seven one-dimensional arrays are allocated: three components of the position, three components of the
velocity, and the charge. The size of each of these arrays is defined in the input file as a multiple of the initial number of particles
per process. In addition six buffers are allocated to communicate the particles to the neighbours. The size of the buffer is defined as
5\% of the total number of particles per process and eventually can be resized if needed. In a typical case, the memory required for
the particles is much larger than the one needed for the fields.  The scheme of the main loop of the code is given by \refAlg{alg:ecsim} and \refFig{fig:scheme}. 
Here the subindex $N$ denotes a magnitude defined on the nodes and $C$ on the centers of the cells.

\begin{algorithm}
    \caption{ECsim main loop}
    \label{alg:ecsim}
    \begin{algorithmic}[1]
        \State \Call{Init}{${\bv}_p, {\br}_p, q_p$}
        \State \Call{Init}{${\bfE}_N^0, {\bfB}_C^0$}
        
        \For{$c \gets 0 \textrm{ to } Cycles$}
           \State $(\widehat{\bJ}, M_{NN'}^{ij}) \gets$ \Call{Moment gathering}{${\bv}_p, {\br}_p$}
           \State $({\bfE}_N^{n+\theta},{\bfB}_C^{n+\theta}) \gets$ \Call{Field solver}{$\widehat{\bJ}, M_{NN'}^{ij}$}
           \State $({\br}_p^{n+1},{\bv}_N^{n+3/2}) \gets$ \Call{Particle mover}{${\bfE}_N^{n+\theta},{\bfB}_C^{n+\theta}$}
        \EndFor
    \end{algorithmic}
\end{algorithm}

\subsection{Moment gathering}
\label{sec:moment_gathering}

It is in this stage where all the magnitudes which are needed for solving the Maxwell equations are computed. This
is where the information from the particles is interpolated to the grid where the field will be calculated.
This task is usually the one with the highest computational cost. 

For each particle we need to compute its contribution to the implicit current and the mass matrices and then interpolate
and accumulate their values in the grid. The scheme of this routine can be seen in \refAlg{alg:moments}

%
%

\begin{algorithm}
    \caption{MOMENT GATHERING. Calculation and interpolation of the implicit current and mass matrices.}
    \label{alg:moments}
    \begin{algorithmic}[1]
        \For{$p \gets 0 \textrm{ to } NP$}
          \State $W_{Np} \gets$ \Call{Weights}{$\br_p^{n+1/2}$, $\br_N$}
          \State $\bfB^n_p \gets$ \Call{Interpolate $N \rightarrow p$ }{$\bfB_N^n$, $W_{Np}$}
          \State $\alpha^n_p = \frac{1}{1+(\beta_s B_p^{n})^2}
                                 \left(\mathbb{I}-\beta_s \mathbb{I} \times \bB_p^n +\beta_s^2 \bB_p^n \bB_p^n \right)$
          \State $\widehat{\bJ}_p = q_p \, \alpha^n_p \bv^n_p$ 

          \For{$N \gets Nodes$}
            \State $\widehat{\bJ}_N = \widehat{\bJ}_N + q_p \bv_p^{n} W_{Np}$

            \For{$N^\prime \gets$ Nodes}
               \State $M_{NN^\prime} = M_{NN^\prime} + W_{Np} W_{N^\prime N} \alpha_p^n$
           \EndFor
         \EndFor
      \EndFor
      \State \Call{Communicate}{$M_{NN^\prime}$,$ \widehat{\bJ}_N$ }
    \end{algorithmic}
\end{algorithm}

For each particle, the first step is to compute the interpolation weights of the particle position with respect to the nodes of the cell of the particle.
Using these weights the magnetic field is interpolated from the nodes to the particle position. These weights will be used later to interpolate the
current and each element of the mass matrix from the particle to the grid. With the magnetic field in the particle position the alpha matrix
 is computed. This matrix will be needed in the mover, but it is not practical to keep them in memory because of the amount of memory required.
Once the implicit current has been computed, it is interpolated to each of the eight nodes of the cell in which the particle is located. For each of these nodes,
we need to compute the contribution of the particle to every component of the mass matrix. 

\begin{figure}[h]
  \center
  \input{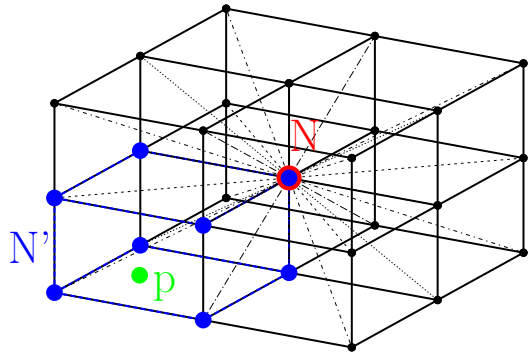}
  \caption{Scheme of the nodes involved in the computation of the mass matrices for a single particle. In red the node where the mass matrix
is computed, in black all the surrounding nodes, in green the particle considered and in blue the nodes which belong to the cell where the particle is.}
  \label{fig:mass01}
\end{figure}

For each node there are 27 mass matrices (in 3D): one for each surrounding node plus that of the node itself. However not all 
these components are affected by every particle, each particle only contributes to the components which correspond to nodes 
in the cell of the particle (see \refFig{fig:mass01} for a better understanding). This is because the interpolation 
function between the particle and the grid is zero for nodes which do not belong to the cell where the particle is located. This means 
that all the mass matrices in which any two nodes of the cell are involved need to be computed. However these mass matrices are 
symmetric with respect to $N N^\prime$: $M_{NN^\prime} = M_{N^\prime N}$, which reduces the number of matrices per node. In particular, 
only 14 matrices per node are needed, the rest can be obtained from the neighbour nodes. With this in mind, the mass matrices are 
stored in memory as a four-dimensional array, the first three indices denote the node $N$ of the grid, and the fourth one the direction 
in which the $N^\prime$ node is.  As we said, for each value of $N$ there are 14 possible directions.

Note that the boundary nodes of each process have only the contribution of the particles which are contained in that process. 
To get the right value there, it is necessary to sum the contribution of all the particles that surround those nodes, which 
means that the values of the boundary nodes from the neighbour processes need to be communicated. This last part is the only one 
in this phase which requires global communications, the rest of the operations are done locally. As will be shown later, this is 
the reason of the very good scaling of this part.  

\subsection{Field solver}

The resolution of the linear system given in \refEqu{EQmaxwell} is done using PETSc. PETSc is a suit of data structures and 
routines specially intended for parallel calculations. Among many other things it provides several linear solvers. In this particular 
case a matrix-free GMRES solver has been used.  Since the method used is matrix-free, the coefficient matrix is not needed. Instead 
the function which computes the product of the coefficient matrix and the solution vector has to be provided. This function computes,
for each node, the parts of \refEqu{EQmaxwell} which depend on $\bfE_N^{n+\theta]}$ and $\bfB_N^{n+\theta]}$:
  \begin{equation}
  \begin{array}{ccc}
    Ax^C &=& \displaystyle \nabla_C \times \bfE_N^{n+\theta} + \frac{1}{c \theta \Delta t} \bfB_C^{n+\theta}\\ \\
    Ax^N &=& \displaystyle \nabla_N \times \bfB_C^{n+\theta} - \frac{1}{c \theta \Delta t} \bfE_N^{n+\theta} - \frac{4\pi}{c} \left(\sum_{N^\prime} M_{NN^\prime} \bE_{N^\prime}^{n+\theta} \right),
  \end{array}
  \label{EQmaxwellimage}
  \end{equation}

The curls are performed with the finite difference approximation. Hence, since the electric field is defined in the nodes, 
its curl is defined in the centers.  And the other way around, the curl of the magnetic field is defined in the nodes. 
This gives us a total of $3 \, (N_x \times N_y \times N_z + C_x \times C_y \times C_z)$ equations, where $C_i$ is the number 
of cells in the $i$ direction and $N_i=C_i+1$ the number of nodes.
On the other hand the independent term is given by the expression:

  \begin{eqnarray}
    b^C &=& \frac{\bfB_C^n }{c \theta \Delta t} \\
    b^N &=& -\frac{\bfE_N^n }{c \theta \Delta t} + \frac{4\pi}{c} \widehat{\bJ}_{g}
  \label{EQmaxwellsource}
  \end{eqnarray}
Thus, the system we need to solve is:
  \begin{equation}
   Ax = b  \rightarrow \left(
  \begin{array}{ccc}
  (Ax^C)_{0,0,0}  \\
  \vdots \\
  (Ax^C)_{C_x,C_y,C_z}  \\
  (Ax^N)_{0,0,0}  \\
  \vdots \\
  (Ax^N)_{N_x,N_y,N_z} 
  \end{array}
  \right)
  =  \left(
  \begin{array}{ccc}
  (b^C)_{0,0,0}  \\
  \vdots \\
  (b^C)_{C_x,C_y,C_z}  \\
  (b^N)_{0,0,0}  \\
  \vdots \\
  (b^N)_{N_x,N_y,N_z} 
  \end{array}
  \right)
  \label{EQmaxwellsource}
  \end{equation}
To compute the curls in \refEqu{EQmaxwellimage} the values of the ghost nodes are needed. Then in every sub-iteration of the solver 
the values in the ghost nodes and centers need to be updated, which requires communication between neighbour processes. 
This causes several communications to be done in this stage, which will provoke a considerable increase in the time of this part 
when the number of processes used increases.

After the system is solved, the values of the fields in the next time step are computed:
\begin{eqnarray}
  \bfE_N^{n+1} &=& (1-\theta) \bfE_N^n + \theta \bfE_N^{n+\theta} \qquad (\mathrm{nodes}) \\
  \bfB_C^{n+1} &=& (1-\theta) \bfB_C^n + \theta \bfB_C^{n+\theta} \qquad (\mathrm{centers})
\end{eqnarray}
At this point the variables \code{Bxc[i][j][k], Byc[i][j][k], Bzc[i][j][k]} contain the value of the magnetic field on the centers 
at time $n+1$ and \code{Bxn[i][j][k], Byn[i][j][k], Bzn[i][j][k]} contain the value of the magnetic field on the nodes at time $n$. 
The values on the node cannot be updated until the particles are moved.

\subsection{Particle mover}

The particle positions and velocities are updated by using the electric field at time $t + \theta$ and the magnetic field 
at time $t$. This ensures that the current created by the particles at time $t + \theta$ is exactly the same as the one 
used in the field equations. For each particle, the same $\alpha$ matrix used in the \textit{Moment gathering} needs to 
be computed. However, due to the large number of particles normally used in simulations these matrices cannot be kept in 
memory and need to be computed again. This means that at this point we have to have in memory the old value of the magnetic 
field on the nodes. After solving the field equations, the new values of the magnetic field are computed in the centers, 
which means that, indeed, the nodes contain the magnetic field at time $t$. Hence, by interpolating these values from the 
nodes to the particles we compute the mass matrix $\alpha$ of each particle. In the same way, the value of the electric 
field at time $t + \theta$ is interpolated from the nodes to the particle position. Once we have the values at the
particle position, the particle velocity is updated and with that the new position is computed (see \refAlg{alg:mover}).

Once all the particles have been moved, those which have left the process domain are communicated to the corresponding process.
These communications are done via \code{MPI\_sendrecv}. For each direction, the particles which have left the domain through that
face are copied into a buffer and then it is sent to the neighbour process in that direction.

\begin{algorithm}
    \caption{PARTICLE MOVER. Update of the particle velocities and positions}
    \label{alg:mover}
    \begin{algorithmic}[1]
        \For{$p \gets 0 \textrm{ to } NP$}
          \State $W_{Np} \gets$ \Call{Weights}{$\br_p^{n+1/2}$, $\br_N$}
          \State $\bfB^n_p \gets$ \Call{Interpolate $N \rightarrow p$ }{$\bfB_N^n$, $W_{Np}$}
          \State $\bfE^{n+\theta}_p \gets$ \Call{Interpolate $N \rightarrow p$ }{$\bfE_N^{n+\theta}$, $W_{Np}$}
          \State $\alpha^n_p = \frac{1}{1+(\beta_s B_p^{n})^2}
                                 \left(\mathbb{I}-\beta_s \mathbb{I} \times \bB_p^n +\beta_s^2 \bB_p^n \bB_p^n \right)$
          \State $\bv_p^{n+1} = 2 \alpha^n_p \left(\bv_p^n + \beta_s \bfE_p^{n+\theta}\right) - \bv_p^n $
          \State $\br_p^{n+3/2} = \br_p^{n+1/2} + \Delta t \bv_p^{n+1}$
        \EndFor
      \State \Call{Communicate}{$\bv_p^{n+1},\br_p^{n+3/2}$}
    \end{algorithmic}
\end{algorithm}

\section{Analysis of performance}
To analyze the performance of the code we have used as a test case an homogeneous plasma  
in 3D formed by electrons and ions with $m_i = 100 m_e$. Their initial thermal speed
is $v_{\mathrm{th}e}/c = 0.1$ and $v_{\mathrm{th}i}/c = 0.01$ in all the directions and
with no drift velocity. The total number of particles in the simulations depends on the 
number of cells, in all the cases 512 electrons and 512 ions per cell are used.
The spatial resolution is $\Delta x = \Delta y = \Delta z = 0.1 d_i$, with $d_i$ the
ion skin depth of the plasma. The time step used is $\Delta t = 0.1 \omega_{pi}^{-1}$ , $\omega_{pi}$ being the ion
plasma frequency; 10 cycles are done. The tolerance of the field solver is set to $10^{-12}$. 
The runtime has been measured with VTune Amplifier \cite{VTune} and with hard coded calls to MPI\_Wtime() function.

We have tested the code in three different systems: 
\begin{itemize}
  \item Marenostrum (Barcelona Supercomputer Center): Two Intel Xeon E5-2670 per node with a combination 
   of Infiniband FDR10 with Gigabit Ethernet to access the hard drives. Used for weak and strong scaling.
  \item SuperMuc (Leibniz-Rechenzentrum): Two Intel Xeon E5-2680 8C Infiniband FDR10. Used for weak and strong
scaling.
  \item DEEP (Juelich Supercomputer Center): Two Intel Xeon E5-2680 with 1 Gigabit-Ethernet and Infiniband (QDR) in a
   three-dimensonal torus (FPGA based). Used for profiling.
\end{itemize}

\begin{figure}[h]
  \center
  \input{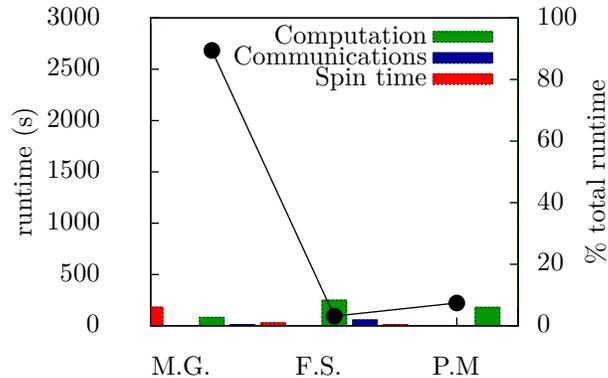}
  \caption{Runtimes of the three main phases of the code with bars (left axis): moment 
           gathering (M.G.), field solver (F.S.) and particle mover (P.M.). In black the 
           percentage of the total time spend in each phase (right scale).
           512 MPI processes, 160 $\times$ 160 $\times$ 160 cells.}
  \label{fig:phases}
\end{figure}

With the help of VTune we have obtained the time that is spent on each part
of the code. In \refFig{fig:phases} we can see the times for the three main phases
of the code. For each one we have measured the computation time (effective CPU time
not including the communications), the communication time (time spent passing 
data from one process to another) and spin time (the wait time while one thread 
is waiting for other threads to finish).

The first thing to point out is that the moment gathering is consuming almost 90\%
of the total runtime of the code. It is in this part where the mass matrices are
calculated, and indeed about 75\% 
of the time of this phase is consumed in their 
calculation. The other two important parts of the code are almost negligible when
compared with the moment gathering. Of course these numbers depend on the number 
of particles per cell. As this value decreases, the computation time of the field solver
becomes more important. However, in a practical case a large number of particles 
per cell (100 at least) is needed to reduce the noise of the simulation.



 
In order to analyse the weak scaling of the code we have used $20 \times 20 \times 20 =
8000 $~cells per MPI process. With that, the code needs around 0.5 Gb 
of memory per process
to store the fields and the particles (including auxiliary arrays and buffers). 


  \begin{figure}[h]
    \center
    \input{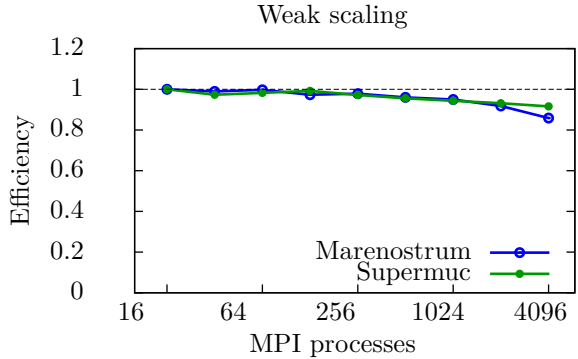}
    \caption{Weak scaling. In ordinates the number of MPI processes and in abscissa the efficiency: runtime 
             with one node (16 MPI processes)  over runtime with the given number of nodes.}
    \label{fig:weak}
  \end{figure}

  \refFig{fig:weak} shows the efficiency of the code on up to 4096 cores in Marenostrum and in Supermuc. The runtime 
  with 16 cores (1 node) has been taken as a reference. In general the code shows a very good scaling in both systems: 
  results from Marenostrum show an efficiency of 86\%  
with 4096 cores while the results from Supermuc are slightly better,
  with an efficiency of 92\% 
 with the maximum number of nodes used here.  
  
  \begin{figure}[h]
    \center
    \input{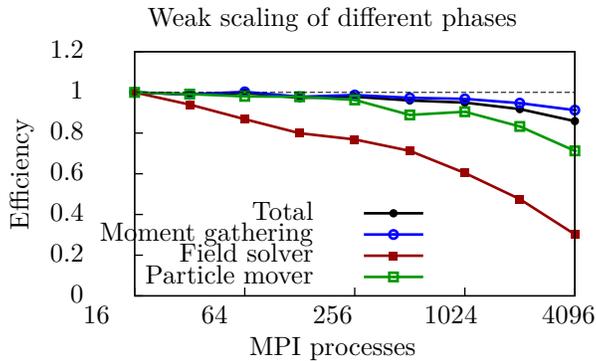}  
    \caption{Weak scaling of different phases of the code. In ordinates the number of MPI processes and 
             in abscissa efficiency: runtime with one node (16 MPI processes)  over runtime with the 
             given number of nodes. Results from Marenostrum.}
    \label{fig:weakphase}
  \end{figure}

  Although the scaling of the code is very good in both systems, it is interesting to analyze which phase of the
  code is responsible for the decrease in the efficiency. In \refFig{fig:weakphase} the efficiency of different
  parts of the code is shown. The first thing we can see is that the trend of the code is ruled by the scaling of the moment
  gathering phase, which scales slightly better than the entire code. The field solver is the phase of the code with 
  worst scaling. However, as we have seen before, this phase consumes very little time when 
  compared with the rest of the code, thus its influence in the total runtime is rather small. Something similar happens
  with the particle mover, which scales better than the field solver but worse than the moment gathering. 

  In general, the cause of the deviation from the ideal scaling are the communications between the processes. In the
  moment gathering phase, once all quantities have been computed in the nodes, it is necessary to communicate them 
  to the rest of the processes, and it is this part the one that causes the loss of efficiency. In the field solver this problem
  becomes more important: in every iteration of the solver the communication of ghost nodes is needed, each process having then to pass data to all the neighbours several times per cycle. In this case the communications are well 
  balanced: each process sends and receives the same amount of data, which helps to reduce the spin time. Finally, in the
  particle mover the communications are required only once after all the particles have been moved. However, the amount of
  data to communicate is not the same for all the processes, as it depends on the number of particles that go in and out of each sub-domain.

  \begin{figure}[h]
    \center
    \includegraphics[width=0.8\columnwidth]{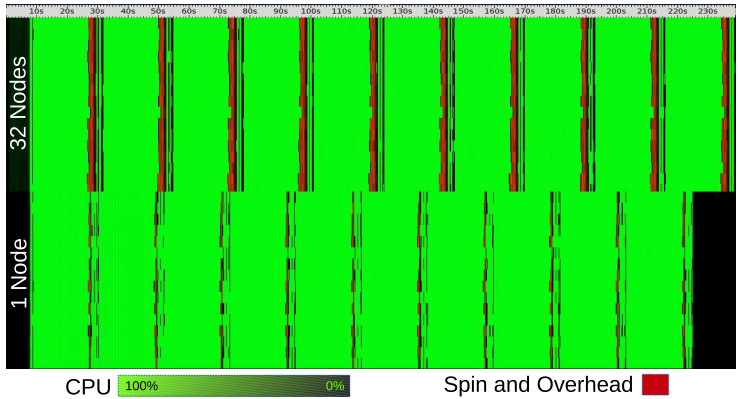}  
    \caption{Weak scaling. Effective CPU time (green-black) and spin time (red) for each thread within one node 
             in two cases: when the code is executed in a single node (bottom) and when it is executed in 32 nodes (top).}
    \label{fig:trazas}
  \end{figure}

  In \refFig{fig:trazas} we can see the effective CPU time and spin time 
  of the 16 threads within one node in two different cases: when the code is executed in a single node (16 MPI processes) and
  when it is executed in 32 nodes (512 MPI processes). In the second case only one node is shown since all of them behave alike.
  These tests have been done in the DEEP system. 
  The first thing we can observe is that the load is very well balanced: all the processors are always doing something, there are no 
  big time 
  gaps where one processor is waiting for the others. 
  However, in both cases we observe that there is a zone of spin time and low CPU efficiency at the end of each cycle. 
  The first band we see in red (wider in the run with 32 nodes) corresponds to the communication of the current and 
  mass matrices after the moment gathering. This is where the highest amount of memory is copied between processes, as can be seen
  in \refFig{fig:phases}. After that, in the field solvers, the ghost values are communicated in the solver sub-iterations, which is 
  seen as the succession of high and low CPU efficiency intervals. At the end of the cycle, the particles are communicated to the 
  right processors and the new cycle starts with the moment gathering, which corresponds to the high-efficiency CPU zone.
  This pattern is repeated 10 times in this example, once per each cycle of the simulation.
  We can observe as well that when 32 nodes are used, the spin time is higher than when only one node is used. 
  It is obvious that in order to improve the scaling of the code we need to optimize or reduce the communications between the processes,
  mainly the ones in the field solver.


  In order to test the strong scaling of the code two sets of calculations have been done. The reason is that
  if we use a case big enough to be executed with 4096 cores, it would takes too long a time to run in only 16 cores.
  And on the other hand, if we choose a case small enough to run in 16 cores, it makes no sense to run it in 4096 cores
  because the computation time would be negligible compared with the communication and synchronization time between
  the processes. In the first case we have used $50 \times 50 \time 100$ cells and it has been run with 16, 32, 64, 128, 256 
  and 512 cores. The second one has $100 \times 100 \time 200$ cells and it has been run with 256, 512, 1024, 2048 and 4096 nodes. 

  \begin{figure}[h]
    \center
    \input{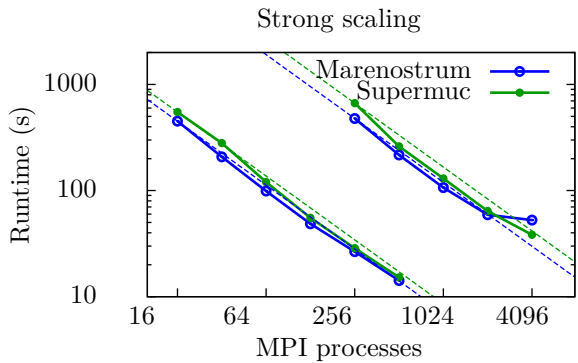}  
    \caption{Strong scaling. Two series of results are shown: one covering between 16 and 512 MPI 
             processes ($50 \times 50 \times 100$ cells) and another between 256 and 4096 MPI processes 
            ($100 \times 200 \times 200$ cells).  In dashed line the ideal scaling.}
    \label{fig:strong}
  \end{figure}

  In \refFig{fig:strong} we can observe the runtimes for the two sets of calculations in Supermuc and Marenostrum.
  We can observe that the efficiency here is much better than in the weak scaling case: Marenostrum shows a scaling almost
  perfect except for the case with 4096 cores, which lasts slightly more than it should. In the case of Supermuc the results
  are even better, the runtimes obtained are shorter than the ideal ones, even with 4096 cores. 
  

\section{Test case: whistler waves}

Whistler waves are  one of the first waves observed in plasmas and they
have been studied for more than 100 years \cite{Stenzel1999}. Whistler waves
are present in magnetized plasmas, and are excited by velocity anisotropy.
Their dispersion relation is given by
eq. \ref{EQwhistler}:
\begin{eqnarray}
	\label{EQwhistler}
  \frac{c^2k^2}{\omega^2} = 1 - \frac{\omega_p^2/\omega^2}{1 - \omega_{c}/\omega} \text{ , } 
\end{eqnarray}
where $\omega_p$ is the plasma frequency, $\omega_c$ the cyclotron frequency,  $c$ the
speed of light and $k$ the parallel component of the wavevector. 

In this section we study a simple case of whistler waves with different spatial
and time resolutions. The results from ECsim are compared with the ones obtained with iPic3D \cite{ipic3d}. 
iPic3D is a semi-implicit particle in cell code which has been used for many years \cite{lapenta2015secondary}\cite{slavik} and 
which is the successor of other previous semi-implicit codes (Celeste3D\cite{Lapenta:2006}, Venus\cite{brackbill-forslund}) in use since the 1980s.  
It uses the Implicit Moment Method (IMM) to remove some of the limitations of the explicit PICs. 
However, in this code energy is not conserved and fields and currents are smoothed in 
order to increase stability. A detailed description of how the smoothing is done in iPic3D is 
shown in Appendix 1.

All the runs are done in two spatial dimensions with a box size of $20\,d_i \; \times \; 1.25\,d_i$. 
The ions have an isotropic Maxwellian velocity distribution with $v_{th,x} = v_{th,y} = v_{th,z} = 0.003 \,c$ 
and the electrons a non-isotropic Maxwellian velocity distribution with $v_{th,x} = 0.01 \, c$, 
$v_{th,y} = v_{th,z} = 0.03 \,c$. The external field is $B_x = B_z = 0$, $B_y = 0.015$, the charge-mass ratio is 100, and 
50 particles per cell have been used. With these parameters, the electron cyclotron frequency is 
$\omega_{ce} = 1.5\,\omega_{pi}$ and the ion cyclotron frequency $\omega_{ce} = 0.015\,\omega_{pi}$.
Here we will focus only on the electrons. Both iPic3D and ECsim use $\theta = 1$ to filter
the high frequency waves.

\subsection{Time resolution analysis}

\begin{figure}[h]
  \center
  \includegraphics[width=0.2\paperwidth]{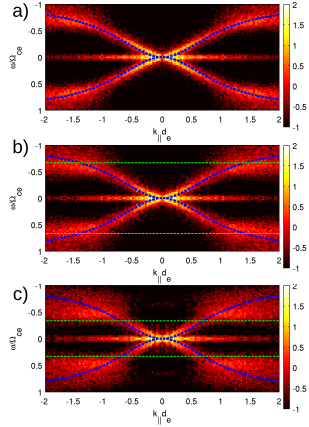}  
  \caption{Omega-k spectra in logarithmic scale obtained from ECsim results for three different values 
           of the time step: 0.5 (a), 1.0 (b) and 2.0 (c). With blue lines the dispersion relation given by 
           equation~\ref{EQwhistler}. In green the maximum value of the frequency resolved in the simulation.}
  \label{fig:whistler_dt}
  \end{figure}

For a given spatial resolution ($512 \times 32$ cells) several time steps have been used (see table \ref{TabDT}).
The total time of each simulation is $160 \, 1/\omega_p$.
With ECsim, the results for the first two cases are essentially the same. Results for the last three cases
are shown in Figure \ref{fig:whistler_dt}. It can be seen that the results from ECsim reproduce the 
theory quite well, even in the cases where the CFL number is larger than 1. Of course, we do not expect to have reliable 
results for frequencies higher than $\omega_{max}$, since they are not resolved in the simulation.
The results from iPic3D for the first two cases (not shown here) are in good agreement with the theory. 
However, for the cases with $\mathrm{d}t=1$ and $\mathrm{d}t=2$, the total energy starts to increase after several iterations 
and the system becomes unstable. It is important to remark here that iPic3D uses smoothing (see Appendix 1) to stabilize 
the system, and even with that it cannot resolve those cases. On the contrary, ECsim does not require any 
smoothing technique.

\begin{table}[h]
	\center
  {
  \begin{tabular}{ccc}
  \hline\hline
	  dt ($1/\omega_p$) & CFL & $\omega_{max}$ ($\omega_{ce}$)\\ 
  \hline
    0.125 & 0.10 & 2.67 \\
    0.500 & 0.38 & 1.33 \\
    1.000 & 0.77 & 0.67 \\
    2.000 & 1.54 & 0.33 \\
  \hline\hline
  \end{tabular} 
  }
  \caption{Values of the time step (first column), CFL number (second column) and maximum frequency resolved
  in the simulations (third column).}
  \label{TabDT}
\end{table}

\subsection{Spatial resolution analysis}

In this case the time step is the same for all runs ($\mathrm{d}t = 0.1$) and the number of cells has been
varied to get different values of the spatial resolution (see table \ref{TabDX}). The CFL number is small enough in
all cases (it is between 0.15 for 1024 cells and 0.019 for 128 cells). In the first two cases iPic3D is stable
even without smoothing due to the high spatial resolution used. In the last two cases, the energy starts to increase, hence the system
will became unstable at some point and in consequence smoothing would be needed for long runs. However, in this particular case, only 1600
cycles are required, and in that time  the total energy only increases about 3 \% 
in the case with 128 cells. This
allows us to run these cases with and without smoothing and compare the results with ECsim.

\begin{table}[h]
	\center
  {
  \begin{tabular}{cccc}
  \hline\hline
	  N. Cells & Spatial resolution & $k_{max}$ $d_e$ & $\Delta x / \lambda_{DE, cold}$ \\
  \hline
    1024 & 0.02 $d_i$ = 0.2 $d_e$ & 5.1 & 0.051 \\
    512  & 0.04 $d_i$ = 0.4 $d_e$ & 2.6 & 0.026 \\
    256  & 0.08 $d_i$ = 0.8 $d_e$ & 1.3 & 0.013 \\
    128  & 0.16 $d_i$ = 1.6 $d_e$ & 0.6 & 0.006 \\
  \hline\hline
  \end{tabular} 
  }
  \caption{Number of cells in the $x$ direction, (first column), grid resolution (second column), maximum wavenumber resolved
  in the simulations (third column) and grid resolution divided by the Debye length of cold electrons ($v_{th} = 0.01$).}  \label{TabDX}
\end{table}

\begin{figure}[h]
  \center
  \includegraphics[width=0.5\paperwidth]{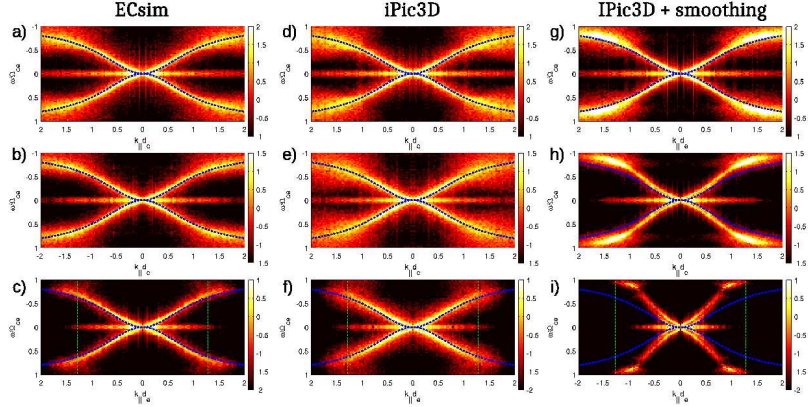}  
  \caption{Omega-k spectra in logarithmic scale obtained from ECsim (a, b and c), iPic3D without smoothing (d, e and f) and iPic3D with smooting (g, h and i). Three different spatial resolutions are shown: 512 cells (a, d and g), 256 cells (b, e and h) and 128 cells (c, f and i).
           With blue lines the dispersion relation given by equation~\ref{EQwhistler}. In green the maximum value of the wavenumber resolved in the simulation.}
  \label{fig:whistler_dx}
  \end{figure}

Figure~\ref{fig:whistler_dx} shows the $k-\omega$ spectra for different spatial resolutions from ECsim and iPic3D with and without smoothing. The first thing we notice is that both ECsim and iPic3D give very accurate results even
with very low spatial resolutions. However, the total energy in iPic3D increases with time, which means that if longer times are needed we will have to use smoothing to stabilize the system. 

When smoothing is used, the effective resolution of the grid decreases due to the average of the electric field, charge and current density (see Appendix 1). This causes the $k-\omega$ to distort even in the range 
where one expects to have reliable results (see figure~ \ref{fig:whistler_dx}).

\subsection{Runtime comparison}
When the same time and spatial resolution are used, ECsim requires more computational time than other semi-implicit PIC codes. This difference is due to the calculation of the mass matrices, which requires
much more time than the moment gathering in other methods (which usually only necessitates the charge and current density and the pressure tensor). On the other hand, the mover in ECsim is simpler than in iPic3D, where
an iterative procedure is used to move the particles. In ECsim the mover has the same scheme as an explicit mover. In a direct comparison between ECsim and iPic3D for the case with 1024 cells 
described before, ECsim needed about 3.2 times more than iPic to run the same case. It is important to keep in mind that the advantage of ECsim is that it can use larger time steps and lower spatial resolution than iPic3D
with the same accuracy (see Figure~\ref{fig:whistler_dx}). 
When one can use a time step one order of magnitude larger and a spatial resolution four times smaller, the fact that its runtime is slighly more expensive than that of other PIC codes is a fair price to pay.

\section{Conclusions}

The three-dimensional, fully electromagnetic, semi implicit particle in Cell code called ECsim has been presented. 
This new code is stable in time, it removes the finite grid instability and it conserves the energy exactly.
When compared with other semi-implicit PICs, the performance tests done show an excellent load balance for the case
proposed here and very good weak and strong scaling.

ECsim has been proven to be capable of tackling situations
where iPic3D presents numerical instabilities. The fact that ECsim does not requires smoothing allows it to 
reproduce physical results with much lower resolution. In these tests we have also found that the smoothing can
dramatically affect the results and can lead to misleading interpretations. Hence, codes which employ smoothing
must be used carefully to always ensure a sufficiently high spatial resolution.

\section*{Acknowledgement}
The research leading to these results has received funding from the European Community's Horizon 2020 (H2020) 
Funding Programme under Grant Agreement n° 754304 (Project „ DEEP-EST“) and has been supported by R\&D Agreement 
between Energy Matter Conversation Corporation (EMC2) and KU Leuven R\&D (Contract \# 2017/771). The computations 
were carried out at the NASA Advanced Supercomputing Facilities (NCCS). This research used resources of the National 
Energy Research Scientific Computing Center, a DOE Office of the Science User Facility supported by the Office of 
Science of the U.S. Department of Energy under Contract No. DE-AC02-05CH11231.

\appendix
\section{Smoothing in iPic3D}

In this appendix we describe the smoothing technique used in iPic3D.
The smoothing partially removes the noise created by the finite number of computational particles and helps to 
stabilize the code. In iPic3D the sources of Maxwell's equations (charge density and implicit current) and the electric
field are smoothed every time step. The smoothing is done by averaging the value of each grid point with its 6 closest neighbours:

\begin{equation}
  \tilde{\rho_g} = \frac{1}{2} \rho_g + \frac{1}{2} \sum_{g\prime} \frac{1}{6} \rho_{g \prime} \text{ , } 
\end{equation}
where $\rho$ is the charge density (the same applies for each component of the electric field and the implicit current), and the sum over $g\prime$ is
extended to the six closest neighbours of g (2 in each direction $x$, $y$ and $z$). This way, the old value at $g$ has the same weight as all 
of the neighbours. For the charge density and the implicit current this procedure is repeated three times for each quantity and nine times for each
component of the electric field.

\section*{References}

\bibliography{bibliography}

\end{document}